# Theoretical results for chemotactic response and drift of *E. coli* in a weak attractant gradient

**Melissa Reneaux§** and **Manoj Gopalakrishnan¶**

§ Department of Physics and Astrophysics, Delhi University, St. Stephen's College, Delhi 110007, India.

¶ Department of Physics, Indian Institute of Technology Madras, Chennai 600036, India.

## Abstract

**The bacterium Escherichia coli (E. coli) moves in its natural environment in a series of straight runs, interrupted by tumbles which cause change of direction. It performs chemotaxis towards chemo-attractants by extending the duration of runs in the direction of the source. When there is a spatial gradient in the attractant concentration, this bias produces a drift velocity directed towards its source, whereas in a uniform concentration, E.coli adapts, almost perfectly in case of methyl aspartate. Recently, microfluidic experiments have measured the drift velocity of E.coli in precisely controlled attractant gradients, but no general theoretical expression for the same exists. With this motivation, we study an analytically soluble model here, based on the Barkai-Leibler model, originally introduced to explain the perfect adaptation. Rigorous mathematical expressions are obtained for the chemotactic response function and the drift velocity in the limit of weak gradients and under the assumption of completely random tumbles. The theoretical predictions compare favorably with experimental results, especially at high concentrations. We further show that the signal transduction network weakens the dependence of the drift on concentration, thus enhancing the range of sensitivity.**

## 1. Introduction

Prokaryotic microorganisms like bacteria have to keep sensing their chemical environment to survive, and are able to adjust their motility in response to changes in it [Bray, 2001]. This ability to bias the motion towards favourable stimuli (*attractants*, eg. Oxygen, nutrients) and away from unfavourable ones (*repellants*, eg. toxins) is referred to as *chemotaxis*. In particular, chemotaxis in the common bacterium *E. coli* has been well-characterized. In a neutral solution, devoid of attractants/repellants, E. coli swims in a zig-zag manner, in a random walk. In the presence of an attractant, however, the walk becomes biased towards the source of the attractant (the reverse happens in the case of a repellant) [Berg, 2003].

How does E. coli bias its motion? Though the abrupt switch in direction during a straight swim might appear to be a stochastic process, it is not really so; rather, this process is regulated by the signal transduction machinery. E. coli senses the attractants through receptor proteins which exist as a single cluster at one pole of the cell. The two main types of receptors are Tar and Tsr. The receptor protein is linked to the protein kinase CheA through the linker protein CheW, and these three are believed to function as a single signaling complex, which exists in active or inactive state, and undergoes stochastic switching between the states [Asakura and Honda, 1984]. In the active state, CheA undergoes auto-phosphorylation and transfers the phosphoryl group to CheY, which diffuses through the cytoplasm and functions as a response regulator [McNab and Koshland, 1972; Koshland, 1977]. CheY binds to the protein FliM at the base of the flagellar motor and increases the rate of switching from the CCW to CW mode of rotation. When one or more flagella thus reverses the sense of rotation, the bacterium tumbles over [Block et. al., 1982; 1983].

The presence of attractants in the solution modifies the chain of events described before. When an attractant molecule binds to a receptor protein, its probability of being active is reduced; therefore, the phosphorylation of CheY is adversely affected, and the frequency of tumbles is reduced. The manifestation of this is that the bacterium tend to spend more time swimming straight without a change of direction. But how does it ensure that it will move towards the source of the attractant? And, how is the motion affected in the presence of a uniform concentration of attractant, as opposed to a spatial gradient? These questions can be answered only by considering another important component of the signal transduction pathway, i.e, methylation and de-methylation processes of the receptor.

The receptor Tar which binds methyl aspartate has five methylation states, with a maximum of four and a minimum of zero methyl groups per receptor. The probability that a receptor-CheA complex is active increases with the methylation level (almost linearly at low attractant concentrations, see Sourjik and Berg, 2002; Kollman et. al., 2005). Methylation of the receptor is accomplished through the protein CheR, and demethylation is done by another protein CheB, in phosphorylated form. Phosphorylated CheB is more efficient at demethylation, and is done by phosphorylated CheA itself, which provides an effective negative feedback in the chain: when attractant binding lowers the mean receptor activity, phosphorylation of CheB is reduced and therefore the mean methylation level goes up, which increases the activity. As a consequence, in the presence of a uniform attractant concentration, a steady state is reached and the system adapts. In the case of E. Coli, the adaptation to methyl aspartate, a common attractant is near-perfect over five orders of magnitude of the background concentrations.

In the presence of spatially varying stimulus, ligand binding and changes in methylation compete with each other, and methylation being a slower process, the receptor activity is consistently reduced over time. Therefore, the tumbles become less frequent whenever the direction of a straight swim (a vector) has a non-zero component along the direction of the attractant gradient. This results in a directionality of motion over several tumbles, and manifests itself as a drift in the direction of the attractant gradient, towards its source. If $L$ is the local chemo-attractant concentration and $\nabla L$ its spatial gradient, then for sufficiently small gradients, we expect that the drift velocity increases proportional to the gradient:

$$v_d \approx \kappa(L)\nabla L \qquad\qquad\qquad\qquad\qquad\qquad\qquad\qquad\qquad\qquad [1]$$

In this article, we present a theoretical calculation of the chemotactic coefficient (or chemotactic sensitivity) $\kappa(L)$ of a single bacterium (i.e. neglecting population effects), for small attractant gradients.

Rivero et. al.(1989) proposed a formula for the chemotactic drift velocity based on ligand-receptor binding kinetics, however, detailed knowledge of the molecular reactions inside the bacterium did not exist at the time. De Gennes (2004) provided an expression which involved the chemotactic response function (response to impulse stimuli), however no detailed modeling was employed to calculate the response function itself. Our attempt here is to utilize the presently accepted model for the intra-cellular reactions, especially the adaptation module and the coupling to the flagellar motor, to compute this response function, and to use it to calculate the drift velocity.

The details of the model are explained in the following section, which is followed by a discussion of the results.

## 2. Model and calculations

### 2.1 Methylation-demethylation reactions

Based on the two-state model of Asakura and Honda [Asakura and Honda, 1984], Barkai and Leibler [Barkai and Leibler, 1997] presented a model to explain the robustness of the perfect adaptation of E.coli against inter-cellular variations in protein concentrations. The model was later extended by Morton-Firth et. al (1999), Mello and Tu (2003) Rao et. al. (2004) and Kollman et. al. (2005). In the presently accepted formulation of the model, the receptor-CheA complex is either active or inactive, and the probability of being in active state depends on the methylation level of the receptor and ligand occupancy. The protein CheR methylates the receptors, and CheB demethylates it. Robust adaptation was achieved in the model through the following assumptions: CheR methylates only receptors in inactive state, while CheB demethylates active receptors. A feedback loop enters the network through the condition that phosphorylated CheB is more efficient at demethylation, and it is (auto-phosphorylated) CheA which provides the phosphoryl groups.

The mathematical description of methylation-demethylation reactions presented here is mostly based on the formalism presented by Emonet and Cluzel (2008) which itself is based on several earlier papers [Barkai and Leibler, 1997; Morton-Firth et. al., 1999; Sourjik and Berg, 2002; Mello and Tu, 2003; Kollman et. al., 2005]. However, it may be noted that our notations differ from that in Emonet and Cluzel, 2008 in many cases.

Let us denote by $X_m$ the fraction of receptor (we consider Tar receptor here, which binds methyl aspartate) complexes with $m$ methyl groups, and let $r$ denote the methylation rate and $b$ denote the demethylation rate of the complex. Let $a_m$ be the fraction of active receptors with $m$ methyl groups. We now assume, in conformity with the assumptions of the Barkai-Leibler model, that CheR binds only to inactive receptors and CheB binds only to active receptors. Then, the dynamical equations for methylation-demethylation reactions can be written in the following form:

$$\frac{dX_0}{dt} = -r(1-a_0)X_0 + ba_1 X_1$$

$$\frac{dX_m}{dt} = r[(1-a_{m-1})X_{m-1} - (1-a_m)X_m] + b[a_{m+1}X_{m+1} - a_m X_m] \quad \text{for } 1 \le m < m_{max} \quad [2]$$

$$\frac{dX_{m_{max}}}{dt} = r(1-a_{m_{max}-1}) - ba_{m_{max}} X_{m_{max}}.$$

where $m_{max} = 4$ for the Tar receptor, and the rates $r$ and $b$ are assumed to have the Michaelis-Menten forms

$$r = \frac{\omega_r R_0}{K_r + A} \quad \text{and} \quad b = \frac{\omega_b B'}{K_b + A^*} \quad [3]$$

where $\omega_r$ and $\omega_b$ are the rates of formation of the final products (methylated and de-methylated receptors) from the inter-mediate complexes of receptor-CheA with CheR and CheB-P respectively, and $B'$ denote the total concentration of CheB-P. $R_0$ is the concentration of CheR, which is assumed to be large, so depletion effects are neglected. $A$ and $A^*$ denote the concentrations of inactive and active receptor-CheA respectively, and are normalized as $A + A^* = A_0$, with $A_0$ being the total concentration of receptor-CheA in the cell (assumed equal to the concentration of CheA).

In order to complete the reaction scheme, we also need to consider the kinetics of CheB-P. This is described by the equation

$$\frac{dB'}{dt} \approx a_P k_P A^*(t) B_0 - d_b \frac{K_b}{K_b + A^*(t)} B'(t) \quad [4]$$

where $a_P$ is the fraction of active receptors in phosphorylated state, $B_0$ is the total concentration of CheB in solution (also assumed large), $k_P$ is the second order association constant for the binding of phosphorylated CheA and CheB and $d_b$ is the dephosphorylation rate of CheB-P in solution. The extra factor $K_b/(K_b + A^*)$ appears because only CheB-P free in solution is dephosphorylated, and not the ones bound to the receptors, and this factor gives the fraction of CheB-P that is free in solution. Since the phosphorylation-dephosphorylation reactions are very fast, one may infer the probability $a_p$ from steady state conditions: $a_p = \omega_p/(k_p B_0 + k'_p Y_0)$, where $Y_0$ is the concentration of free CheY in solution, $k'_p$ is the second order rate constant for its binding to phosphorylated CheA, and $\omega_p$ is the rate of auto-phosphorylation of active CheA.

Finally, the concentration of active receptor-CheA complexes is given by

$$A^* = A_0 \sum_{m=0}^{m_{max}} a_m X_m \qquad [5]$$

Equations [2-5] form a complete set of equations, which predicts the kinetics of evolution of receptor activity in terms of the activation probabilities $a_m$. The activation probabilities depend on whether the receptor is liganded or not. It has been reported that, for small ligand concentrations, this probability may be well approximated by the formula [Sourjik and Berg, 2002; Kollman et. al., 2005]

$$a_m = a_m^0 \left( \frac{K_m^H}{K_m^H + L^H} \right) \qquad \text{for } 1 \leq m < m_{max}, \qquad [6a]$$

where $a_m^0$ is the response amplitude, $K_m$ is the dissociation constant for the ligand-receptor binding reaction when the receptor has m methyl groups. $H$ is the Hill coefficient, which is experimentally measured to be 1.2. Table I lists the experimentally determined values of $K_m$ for different methylation levels [Kollman et. al., 2005].

## 2.2 Reduction to three-state models

As shown by Mello and Tu (2003), perfect adaptation is achieved in the Barkai-Leibler model if the boundary values of activation probability remain fixed at two extreme values at all concentrations: i.e., $a_0 = 0$ and $a_{m_{max}} = 1$ independent of $L$. Keeping this constraint intact, the model can be considerably simplified if the number of methylation levels are effectively reduced to three: (i) $M = 0$, which is always inactive (ii) $M = 1$, whose activity depends on $L$ through a relation of the form as in Eq.6, and (iii) $M = 2$, which is always active. In this reduced model, we therefore have $a_0 = 0$, $a_2 = 1$, while $a_1$ depends on $L$ through the following equation:

$$a_1(L) = a_1^0 \left( \frac{K_D^H}{K_D^H + L^H} \right) \qquad [6b]$$

where $K_D$ is the dissociation constant for the lone intermediate methylation state, which needs to be chosen from among the different $K_m$. The precise relation between the two quantities is explained below.

A close look at the experimental values for the dissociation constants and activation probabilities at various methylation levels of the Tar-receptor (Table I) shows that the three-state approximation works well over a wide range of attractant concentration. Consider first, the range $20\mu M \ll L \ll 1500\mu M$. In this range, the methylation states 3 and 4 have activation probability $a_m$ very close to 1, from Eq. 6a, while for states 0 and 1, $a_m \approx 0$ in the entire range. Therefore, this regime can be described by a three-state model as described above with $K_D = 150 \mu M$ and $a_1^{(0)} = 1/2$, both of which correspond to methylation level $m = 2$ in the original system. We refer to this regime of attractant concentration as L1 in later analysis.

A second, slightly different three-state model may be used to study the regime $150 \mu M \ll L \ll 60 mM$. In this case, the original methylation states $m = 0,1,2$ can be collapsed to a single $M = 0$ state with $a_0 = 0$, while methylation state $m = 3$ becomes the new intermediate state $M = 1$ with $K_D = 1500 \mu M$ and $a_1^{(0)} = 3/4$. This regime will be referred to as L2 henceforth wherever required.

Between these two versions of the three-state model, a wide range of attractant concentrations can be studied. The three-state approximation considerably simplifies the mathematical equations and permits a completely analytical solution to the problem of computing the drift velocity.

**2.3 Steady state in uniform attractant concentration**

In the reduced Barkai-Leibler model, we have three receptor populations $X_M$ with $M = 0,1,2$, which satisfy the following dynamical equations, which follow from Eq.2.

$$\frac{dX_0}{dt} = -rX_0 + ba_1(L)X_1$$
$$\frac{dX_2}{dt} = r[1 - a_1(L)]X_1 - bX_2$$
[7]

By normalization, $X_1 = 1 - X_0 - X_2$. Let us now assume that the bacterium is placed in a spatially uniform attractant concentration $L$ for sufficiently long time so that the system reaches a steady state and all $X_M(t) = X_M^{(0)}$, for which explicit expressions are obtained from Eq.7 by putting time derivatives to zero.

$$X_0^{(0)} = \frac{a_1(L)}{a_1(L) + c + c^2[1 - a_1(L)]}; X_1^{(0)} = \frac{c}{a_1(L)} X_0^{(0)}; X_2^{(0)} = 1 - X_0^{(0)} - X_2^{(0)},$$
[8a]

where $c = r^{(0)}/b^{(0)}$, with

$$r^{(0)} = \frac{\omega_r R_0}{K_r + A^{(0)}}; b^{(0)} = \frac{\omega_b B'^{(0)}}{K_b + A^{*(0)}}; B'^{(0)} = a_P \frac{k_P}{d_b} A^{*(0)} \left(1 + \frac{A^{*(0)}}{K_b}\right)$$
[8b]

and the steady state active kinase fraction given by

$$a^{*(0)} = \frac{A^*}{A_0} = a_1(L)X_1^{(0)} + X_2^{(0)}$$
[8c]

Eq.8a-8c may be solved self-consistently now to determine the active kinase fraction, which, after using Eq.8a and 8c, turns out to be independent of $a_1(L)$. This, of course, is the requirement of perfect adaptation which follows by construction in the Barkai-Leibler model.

## 2.4 Attractant gradient as a weak perturbation

We will now `switch on' a spatial gradient $\nabla L$ in the concentration at time $t = 0$. In the course of its motion, therefore, the bacterium now experiences an attractant concentration that changes with time, a direct consequence of which is that $a_1$ is now time-dependent, i.e.,

$$a_1(L) \to a_1(L) + \delta a_1(t), \text{ where } \delta a_1(t) = \frac{\partial a_1}{\partial L} \delta L(t) \tag{9}$$

where $\delta L(t) = \int_0^t d\tau \dot{L}(\tau)$, with $\dot{L}(t)$ being the *rate of change of the attractant concentration experienced by the bacterium along its path,* which will be discussed in detail in the next section.

The change in the activation probability induces corresponding small changes in the other quantities; in particular, we have, at a later time $t > 0$,

$$X_M(t) = X_M^{(0)} + \delta X_M(t) \qquad 0 \le M \le 2 \tag{10}$$

which, from Eq.5, produces a corresponding change in the fraction of active kinases given by

$$\delta a^* = X_1^{(0)} \delta a_1 - a_1(L) \delta X_0 + [1 - a_1(L)] \delta X_2 \quad . \tag{11}$$

From Eq.3 and 4, the quantities $\delta r$ and $\delta b$ depend on $\delta a^*$ as follows:

$$\delta r = \theta_1 \delta a^* ; \delta b = -\theta_2 \delta a^* + \theta_3 \int_0^t e^{-\beta(t-\tau)} \delta a^*(\tau) d\tau \quad , \tag{12}$$

with the coefficients given by

$$\theta_1 = \frac{w_r R_0 A_0}{\left(K_r + A_0(1-a^{*(0)})\right)^2} ; \theta_2 = \frac{w_b B'^{(0)} A_0}{\left(K_b + A_0 a^{*(0)}\right)^2} ; \theta_3 = \frac{w_b \alpha A_0}{(K_b + A_0 a^{*(0)})} \tag{13a}$$

and

$$\alpha = k_p a_p B_0 \left[ 1 + \frac{A^{*(0)}}{K_b + A^{*(0)}} \right]. \tag{13b}$$

The time integral in the second part of Eq.12 is an integral feedback term, which provides a finite memory to the system. This is seen by considering the time evolution of the perturbations in the fraction of receptor-CheA complexes in different methylation states, which, following Eq.7, satisfy the equations

$$\frac{d\delta X_0}{dt} = -\left[r^{(0)} + a_1(L)b^{(0)}\right]\delta X_0 - a_1(L)b^{(0)}\delta X_2 - X_0^{(0)}\delta r + b^{(0)}X_1^{(0)}\delta a_1(t) + a_1(L)X_1^{(0)}\delta b$$

$$\frac{d\delta X_2}{dt} = -r^{(0)}[1 - a_1(L)](\delta X_0 + \delta X_2) - r^{(0)}X_1^{(0)}\delta a_1(t) + [1 - a_1(L)]X_1^{(0)}\delta r - b^{(0)}\delta X_2 - X_2^{(0)}\delta b$$

[14]

with the coefficients given in Eq.8a-c. To obtain explicit solutions analytically, it is helpful to re-express these equations using Laplace transforms: $\tilde{f}(s) = \int_0^\infty f(t)e^{-st}dt$. After some algebra, using Eq.9-14 (see details in Appendix A), we find that the change in net kinase activity is given by

$$\delta \tilde{a}^*(s) = \tilde{\chi}_a(s)\delta \tilde{a}_1(s), \qquad [15]$$

where $\tilde{\chi}_a(s)$ is the Laplace transform of the *linear response function* $\chi_a(t)$ for kinase activity, defined through the usual integral relation $\delta a^*(t) = \int_0^t \chi_a(t-\tau)\delta a_1(\tau)d\tau$, equivalent to Eq.15. The explicit form of $\tilde{\chi}_a(s)$ is given in Appendix A. It may also be verified, after some calculations, that $\tilde{\chi}_a(0) = 0$, i.e., the area under the response function curve vanishes, which is a requirement for the perfect adaptation property.

Appendix B presents, as a special case, a calculation for $\chi_a(t)$ in the limit of small attractant concentrations.

**2.5 Kinase activity to tumbling regulation**

As explained in the introduction, changes in the activity of the kinase CheA are directly coupled to the flagellar motors, because CheA also phosphorylates CheY (in addition to CheB), and the phosphorylated CheY binds to the base of the flagellar motors and induces a change in the direction of rotation. It has also been shown experimentally that the flagellar motor is ultra-sensitive with respect to CheY-P; the probability of rotating clockwise, the CW bias, can be well approximated by a Hill-type expression with an exponent close to 10 [Cluzel et. al., 2000]. Accordingly, we assume the following form for the rate of CCW-CW switching of the motor:

$$R_- \propto Y'^{H^*} \text{ with } H^* \approx 10 \qquad [16]$$

where $Y'$ is the concentration of CheY-P in solution. CheY-P is dephosphorylated by CheZ in solution, with a rate $k_z$. The change of $Y'$ after switching on the gradient satisfies the equation

$$\frac{d\delta Y'}{dt} = k'_p Y_0 a_p A_0 \delta a^*(t) - k_z \delta Y'(t) \quad , \qquad [17a]$$

while the steady state concentration prior to introducing the gradient is given by

$$Y'^{(0)} = \frac{k'_p Y_0 a_p}{k_z} A^{*(0)} \quad . \qquad [17b]$$

Eq.17a has the formal solution $\delta Y'(t) = k'_p Y_0 a_p A_0 \int_0^t e^{-k_z(t-t')} \delta a^*(t')dt'$, which, after Laplace-transformation, takes the form

$$\delta \widetilde{Y}'(s) = \frac{k'_p Y_0 a_P A_0}{(s+k_z)} \delta \widetilde{a}^*(s) \qquad [17c]$$

From Eq.16, the corresponding fractional change in the CCW-CW switch rate is given by

$$\frac{\delta \widetilde{R}_-(s)}{R_-^{(0)}} = H^* \frac{\delta \widetilde{Y}'(s)}{Y'^{(0)}} = \widetilde{\chi}_R(s) \delta \widetilde{a}_1(s) \quad, \qquad [18]$$

where $R_-^{(0)}$ is the switch rate in steady state, in the absence of attractant and $\widetilde{\chi}_R(s)$ is the linear response function of the switch rate to the change in attractant concentration, which, using Eq.15,17, 18 and 19, is related to $\widetilde{\chi}_a(s)$ through

$$\widetilde{\chi}_R(s) = \frac{H^*}{a^{*(0)}} \frac{k_z}{(s+k_z)} \widetilde{\chi}_a(s) \qquad [19]$$

where $a^{*(0)}$ is the fraction of active kinases in steady state, which is independent of the attractant concentration $L$. This, again, is simply a requirement for perfect adaptation and is ensured by construction in the Barkai-Leibler model.

## 2.6 Relating the response function to drift velocity

Without loss of generality, we now assume that the attractant concentration gradient is directed along the z-axis with a maximum at origin, i.e., $\nabla L = -L_z \hat{z}$, with $L_z$ giving the absolute magnitude of the gradient. Let $\theta_n$ be the angle made by the direction of motion of the bacterium on the positive z-axis, after the $n$'th tumbling event. Between n'th and n+1'th tumbling events, therefore, the bacterium experiences an effective rate of change in the attractant concentration given by

$$\dot{L}_n^{eff} = -vL_z \cos\theta_n, \qquad [19]$$

where $v$ is the velocity of smooth swimming of the bacterium between tumbles.

The probability that the (n+1)'th tumble takes place during the time interval $[\tau_{n+1} : \tau_{n+1} + \Delta\tau_{n+1}]$, with the last tumble having taken place at $t = \tau_n$, is given by $P(\tau_n, \tau_{n+1})\Delta\tau_{n+1}$, where

$$P(\tau_n, \tau_{n+1}) = R_-(\tau_{n+1}) \exp\left(-\int_{\tau_n}^{\tau_{n+1}} R_-(\tau')d\tau'\right). \qquad [20]$$

The mean run interval between the n'th and (n+1)'th tumble is

$$\langle \tau \rangle_n = \int_0^\infty P(\tau_n, \tau_n + T) T dT, \qquad [21]$$

and the change in it, corresponding to the gradient-induced change in the switch-rate is given by

$$\delta \langle \tau \rangle_n = \int_0^\infty TdT \delta R_-(\tau_n + T)e^{-R_-^{(0)}T} - R_-^{(0)} \int_0^\infty TdTe^{-R_-^{(0)}T} \int_{\tau_n}^{\tau_n+T} \delta R_-(\tau)d\tau \qquad [22]$$

From Eq.18,

$$\delta R_-(t) = R_-^{(0)} \int_0^t \chi_R(t-t')\delta a_1(t')dt', \qquad [23]$$

which may be written in the following expanded form, for $t > \tau_n$:

$$\delta R_-(t) = R_-^{(0)}\left[I_t(t,\tau_n)\cos\theta_n + I_t(\tau_n,\tau_{n-1})\cos\theta_{n-1} + \ldots\ldots\right] \qquad [24]$$

where the integrals $I_t(\tau_1,\tau_2) = -vL_z(\partial a_1/\partial L)\int_{\tau_1}^{\tau_2} \chi_R(t-\tau')\tau'd\tau'$ using Eq.9,19 and 23.

If the directions of smooth motion (runs) between successive tumbles are completely uncorrelated, then only the first term in Eq.24 ultimately matters in the determination of the drift velocity (see Eq.26 later). In reality, bacterial run-and-tumble motion exhibits persistence of direction, i.e., the forward run after a tumble is more likely to have a positive component along an axis parallel to the swim-direction before the tumble. Nevertheless, for the sake of simplicity, we proceed on the assumption that the run directions before and after a tumble are entirely uncorrelated. In this case, after each tumble, the organism completely loses memory of the previous run, and we can put $\tau_n = 0$ in Eq.22 without any loss of generality. After a few algebraic calculations, it can be shown from Eq.22 that

$$\delta\langle\tau\rangle_n = -\sigma \frac{v}{\left(R_-^{(0)}\right)^2} \frac{\partial a_1}{\partial L} L_z \cos\theta_n \qquad [25a]$$

where

$$\sigma = \left(\frac{k_Z}{k_Z + R_-^{(0)}}\right)\frac{H^*}{a^{*(0)}} \tilde{\chi}_a(R_-^{(0)}) \qquad [25b]$$

is a dimensionless constant. The mean displacement of the bacterium in the $-z$ direction (towards the attractant source) after $N$ tumbles is given by

$$\langle\Delta x\rangle_N = \sum_{k=0}^N v\langle \tau_k \cos\theta_k \rangle \qquad, \qquad [26]$$

where the averaging needs to be carried out, first over $\tau_k$ for fixed $\theta_k$, and then over $\theta_k$. After using Eq. 25a and carrying out the final angular averaging, we obtain

$$\langle \Delta x \rangle_N = -\frac{\sigma}{2} \frac{v^2}{\left(R_-^{(0)}\right)^2} \frac{\partial a_1}{\partial L} L_z N \quad , \tag{27}$$

where, the result $\langle \cos^2 \theta_k \rangle = 1/2$ has been used. The mean number of tumbles over a time interval $T$ is given by $\bar{N} = R_-^{(0)} T$. Averaging over $N$ for fixed $T$ in Eq.27 therefore yields the drift speed:

$$v_d = -\frac{\langle \Delta x \rangle_T}{T} = -\frac{\sigma}{2} \frac{v^2}{R_-^{(0)}} \frac{\partial a_1}{\partial L} L_z \tag{28}$$

After comparison with Eq.1, we finally arrive at an expression for the chemotactic coefficient $\kappa(L)$ for small gradients, within our reduced model:

$$\kappa(L) = v \frac{\sigma}{2} l_{run} \frac{\partial a_1}{\partial L} \tag{29}$$

where $l_{run} = v / R_-^{(0)}$ is the 'un-perturbed' mean run length and $\sigma$ is defined in Eq.25b.

## 3. Results

Expressions similar to Eq.29 for the chemotactic coefficient have been obtained by previous authors. De Gennes (2004) obtained an expression for $\kappa$ proportional to the Laplace transform of the response function $\chi_R$ (in our notation) at the tumbling frequency, but the details of the response function were not obtained from the network. Locsei (2007) extended De Gennes' calculation to include persistence of direction and rotational diffusion, but again, the detailed mathematical form of the response function was chosen only to resemble the experimentally known bi-lobe form [Block, Segall and Berg, 1982,1983; Segall, Block and Berg, 1986], and not through any consideration of the underlying reactions. To our knowledge, Eq.29 is the first theoretical expression where the drift velocity has been computed in terms of a linear response function directly derived from the biochemical network. The dimensionless number $\sigma$ includes all the biochemistry from methylation and phosphorylation reactions, while $\partial a_1 / \partial L$ includes details of ligand binding, kinase activation and receptor cooperativity.

Eq.29 may now be used to compute the drift velocity of a population of bacteria swimming under the influence of an attractant gradient. A list of experimentally measured/estimated parameters used in the model is given in Table II. In addition, we used a swim speed $v \approx 20 \mu$m/s and tumbling frequency $R_-^{(0)} = 0.5 s^{-1}$, corresponding to a run length $l_{run} \approx 40 \mu$m.

The Laplace transform of the response function given in Eq.33a could be explicitly inverted in the limit of extremely small ligand concentrations, theoretically, the $L \to 0$ limit. The details of the calculation are given in Appendix II, along with the final expression. It turns out, however, with the list of parameters in Emonet and Cluzel (2008), the time scale of decay of the response function is ~ 20 s, four times larger than the experimental value [Block, Segall and Berg, 1982;

Segall, Block and Berg, 1986]. However, increasing the Che-B concentration to $2.0\,\mu M$ [Morton-Firth et. al., 1999; Rao et. al., 2004] produced a curve that decays in slightly over 5 seconds, but with a second, non-pronounced peak at around 12 seconds [Fig.1]. This second peak/over-shoot was not observed in the experiments and is probably an artifact of the model. The CheY-P dephosphorylation rate was also reduced to $10 s^{-1}$ for better agreement of the shape of the curve with experiments, although the effect of this change in the drift velocity is very small (see Eq.25b) since $R_-^{(0)} << k_Z$.

Table III gives a comparison between the drift velocity obtained from our model and experimental values measured by Ahmed and Stocker [Ahmed and Stocker, 2008] recently from microfluidic experiments with the chemo-attractant $\alpha$-methyl-aspartate, for various values of $L$ and $L_z$. As explained in the beginning, we use $K_D = 150\,\mu M$ and $a_1^{(0)} = 1/2$ for $L << 150\,\mu M$ (which applies to the first two measurements) and $K_D = 1.5 mM$ and $a_1^{(0)} = 3/4$ in the concentration range $150\,\mu M << L << 60 mM$ (which applies to the third and fourth measurements quoted in Table II). For the parameters we used, the steady state active kinase fraction was found numerically (using Eq.8a and 8c) to be $a^{*(0)} \approx 0.08$.

In general, we observe that the theoretical calculation underestimates the drift velocity by a factor of 4 to 6. An obvious reason may be our assumption of completely uncorrelated tumbles: as explained in the previous section, we used an approximation whereby a tumble will completely erase the memory of the direction of the run preceding it, but for E. coli, it is now well-established that this is not the case. It has indeed been shown in a previous paper [Locsei, 2007] that persistence of direction may enhance the drift velocity up to a factor of 2 compared to completely random tumbles. However, the possibility that the model itself may be missing some essential ingredients cannot be ruled out. In spite of this, we find it gratifying that the model manages to reproduce the observed drift velocities at least order-of-magnitude-wise, and follows the same qualitative trends (increasing/decreasing) as in experiments.

### 3.1 Logarithmic sensing?

Does gradient sensing by E. Coli follow Weber's law for sensory systems, i.e., does E.coli measure the gradient of $\log(L)$, rather than $L$ itself? It is known that chemotactic sensitivity of E.coli extends over several decades of attractant concentrations and it is, therefore, likely that the sensing takes place on a logarithmic scale. A set of recent microfluidic experiments [Kalinin et. al., 2009] has provided evidence that the drift velocity increases roughly proportional to $\nabla L / L$ over a range of $L$. However, logarithmic sensing has never been rigorously shown to result from any theoretical model so far.

From Eq.1, the essential requirement for logarithmic sensing is that $\kappa(L)$ decreases inversely proportional to $L$ over the range of $L$ in question. From Eq.25b and 29, two factors are seen to depend on $L$, $\tilde{\chi}_R(R_-^{(0)})$ and $\partial a_1 / \partial L$. From Eq.6b, we find that

$$\frac{\partial a_1}{\partial L} = \frac{H}{K_D} \frac{l^{H-1}}{\left(1 + l^H\right)^2} \quad \text{where} \quad l = \frac{L}{K_D} \quad \text{and } H \approx 1.2. \qquad [30]$$

which stays constant when $L \ll K_D$ but decays as $L^{-2.4}$ for much larger $L$. Clearly, the possibility of logarithmic sensing depends crucially on whether $\tilde{\chi}_R(R_-^{(0)})$ increases with $L$, and how.

Fig.2 shows the variation of $\tilde{\chi}_R(R_-^{(0)})$ against the dimensionless concentration $L/K_D$ on a logarithmic scale. It is observed that $\tilde{\chi}_R(R_-^{(0)})$ increases roughly proportional to $\sqrt{L}$ (see the fitting curve) up to almost $L = 10K_D$. Although this rise is too weak to make $\kappa(L)$ decay strictly inversely proportional to $L$, it could potentially make the decay weaker in comparison to the asymptotic $L^{-2.4}$ mode, in an intermediate regime.

Fig.3 shows the variation of $\kappa(L)$ against $L/K_D$ on a logarithmic scale, in the L2 range, with a fitting line proportional to $L^{-1}$ shown for comparison. We observe that the curve approximately follows a $L^{-1}$ decay in a small regime, between 1.5 and 4.5 in the dimensionless concentration $L/K_D$, which corresponds to a narrow range of attractant concentration: $2.25 mM \leq L \leq 6.75 mM$. No such logarithmic regime could be identified in L1. It appears that the regime of applicability of logarithmic sensing, as predicted in our model is limited when compared to the 2-3 orders in magnitude in concentration as suggested by experiments. Choosing a different set of parameters (eg. having $B_0 = 0.28 \mu M$ as chosen in Cluzel et. al., 2008) could potentially extend this regime to the left, as we have observed, but at the risk of affecting other quantities, eg., the time scale of adaptation.

Fig.4 shows the variation of $\kappa(L)$ against $L/K_D$ on a normal scale. We see that the chemotactic sensitivity is maximized at around $L \approx 0.5 K_D$, and not at $L \approx K_D$ as one might have naively expected based on ligand-binding kinetics. Clearly, the entire signal transduction network in E.coli, in particular, the methylation-demethylation module with its integral negative feedback, is important in determining the transport characteristics as well, in addition to adaptation.

## 4. Discussion

In this paper, we have presented a comprehensive phenomenological model to describe quantitatively the chemotactic response of E.coli to changes in attractant concentration in the background. For the methylation-adaptation part, we used the well-known Barkai-Leibler model, wherein it was shown that the experimentally observed near-perfect adaptation of E.coli to Methyl Aspartate can be reproduced in a robust manner, by using a few assumptions. This `methylation-adaptation module' is then connected to the other end of the network, i.e., the CheY/CheZ-flagellar motor module through the `ultra-sensitive' response relation given by Eq.16, supported by experiments.

We derived the chemotactic response function from our model, which has the same bi-lobe form as that obtained by Segall, Block and Berg (1986), and zero total area (which is a direct consequence of the assumptions in Barkai-Leibler model). The response function is then used to compute the drift velocity of a population of bacteria in response to a small gradient in the attractant concentration, under the assumption that the directions of adjacent runs separated by a tumble are completely uncorrelated. The final expression for the drift velocity has three parts: (a) the derivative of the activation probability of the receptor-kinase complex with respect to the attractant concentration (b) a dimensionless parameter which includes all contributions from the biochemical network between ligand binding and flagellar rotation, and (c) the run-length and tumble rate to take care of dimensions.

As a test of our model, we compared the results for drift velocity predicted by the model with recent microfluidic experimental results of Ahmed and Stocker (2008), for various mean concentrations and gradients. In general, the theory underestimates the drift velocity. For small concentrations, our results are smaller by a factor of 6, whereas for larger concentrations, the discrepancy is reduced. While a precise match between theory and experiments might be unrealistic to expect in a complex system like E.coli, we identified two possible reasons for the discrepancy: our assumption of uncorrelated run-directions before and after a tumble and the assumption of small attractant gradient. While the first limitation can be taken care of within our model, by extending the formalism, we undertook a qualitative check of the effect in the following way: since the effect of persistence of the run-direction after a tumble effectively `increases' the length of a smooth run, we checked how the tumble-rate dependent part of the drift velocity depends on this rate. In Fig.5, we plot the factor $\tilde{\chi}_a(R_-^{(0)})/R_-^{(0)}$ against $R_-^{(0)}$. It is observed that smaller tumble rate increases this factor, and consequently the drift velocity. For high concentrations $L \geq K_D$, reducing the tumble rate by ½ increases the drift velocity by a factor of 2 or more. The second limitation, i.e., the assumption of negligibly small gradients is more difficult to handle within the present model, and a new formalism may be necessary.

The last issue we looked at is the possibility of logarithmic chemosensing by E.coli, which has been suggested some time ago, and which appears to be corroborated by recent experiments. Our model has only been partially successful in reproducing this behavior, in a limited range of concentration. From a purely modeling point of view, tuning the numerical values of various parameters could potentially extend the regime of validity of log-sensing, but could affect other results, like the time scale of adaptation and drift velocity. Therefore, the broader questions that need to be addressed are centered around optimization: criteria for optimization of response, foraging, noise reduction etc. It is therefore clear that much more needs to be done in understanding the efficiency of chemosensing of E.coli, for which the drift velocity is only one parameter, though possibly the most easily measurable one.

**Acknowledgements**

MR would like to acknowledge the Harish-Chandra Research Institute, Allahabad and Indian Institute of Technology, Madras for summer research fellowships, which made this work possible. MG would like to thank Mahesh Tirumkudulu (IIT Bombay) for sharing details of experiments and many fruitful discussions.

**Appendix A: Details of mathematical calculations for the response function**

Starting from Eq.14, the Laplace transforms $\delta \widetilde{X}_0(s)$ and $\delta \widetilde{X}_2(s)$ can be seen to satisfy the set of two couple equations

$$(s + R_1)\delta \widetilde{X}_0(s) = -b^{(0)} a_1^{(0)} \delta \widetilde{X}_2(s) + b^{(0)} X_1^{(0)} \delta \widetilde{a}_1(s) - X_0^{(0)} \delta \widetilde{r}(s) + a_1^{(0)} X_1^{(0)} \delta \widetilde{b}(s) \qquad \text{[A.1a]}$$

$$(s + R_2)\delta \widetilde{X}_2(s) = -r^{(0)}(1 - a_1^{(0)}) \delta \widetilde{X}_0(s) - r^{(0)} X_1^{(0)} \delta \widetilde{a}_1(s) - X_2^{(0)} \delta \widetilde{b}(s) + (1 - a_1^{(0)}) X_1^{(0)} \delta \widetilde{r}(s) \qquad \text{[A.1b]}$$

where $R_1 = r^{(0)} + a_1^{(0)} b^{(0)}$ and $R_2 = r^{(0)}(1 - a_1^{(0)}) + b^{(0)}$.

The solution to Eq. A.1a and A.1b may be written in the form

$$\delta \widetilde{X}_0 = \frac{\Delta_1 - h_1 \Delta_2}{1 - h_1 h_2} \; ; \; \delta \widetilde{X}_2 = \frac{\Delta_2 - h_2 \Delta_1}{1 - h_1 h_2} \qquad \text{[A.2a]}$$

where

$$\Delta_1 = \frac{b^{(0)} X_1^{(0)} \delta \widetilde{a}_1 + a_1^{(0)} X_1^{(0)} \delta \widetilde{b} - X_0^{(0)} \delta \widetilde{r}}{s + R_1} \; ; \; \Delta_2 = \frac{(1 - a_1^{(0)}) X_1^{(0)} \delta \widetilde{r} - X_2^{(0)} \delta \widetilde{b} - r^{(0)} X_1^{(0)} \delta \widetilde{a}_1}{s + R_2} \qquad \text{[A.2b]}$$

and

$$h_1 = \frac{a_1^{(0)} b^{(0)}}{s + R_1} \; ; \; h_2 = \frac{r^{(0)}(1 - a_1^{(0)})}{s + R_2} \qquad \text{[A.2c]}$$

From Eq.12, we have

$$\delta \widetilde{r}(s) = \theta_1 \delta \widetilde{a}^*(s) \text{ and } \delta \widetilde{b}(s) = \left( \frac{\theta_3}{s + \beta} - \theta_2 \right) \delta \widetilde{a}^*(s) \qquad \text{[A.3]}$$

Substituting Eq.A.3 into Eq.A.2b, and using Eq.A.2b in Eq.A.2a, and finally using Eq.11 leads us to the linear response relation between $\delta \widetilde{a}^*(s)$ and $\delta \widetilde{a}_1(s)$ in Eq.15, with $\widetilde{\chi}_a(s)$ given by

$$\tilde{\chi}_a(s) = X_1^{(0)} \left[ \frac{(s+R_1)(s+R_2) - b^{(0)}(\alpha_2 + h_2\alpha_1)(s+R_2) - r^{(0)}(\alpha_1 + h_1\alpha_2)(s+R_1)}{(s+R_1)(s+R_2) + \Phi_1[\alpha_2 + h_2\alpha_1](s+R_2) - \Phi_2[\alpha_1 + h_1\alpha_2](s+R_1)} \right],$$ [A.4a]

where

$$\alpha_1 = \frac{1-a_1^{(0)}}{1-h_1 h_2}; \alpha_2 = \frac{a_1^{(0)}}{1-h_1 h_2}$$ [A.4b]

and

$$\Phi_1 = a_1^{(0)} X_1^{(0)} \left( \frac{\theta_3}{(s+\beta)} - \theta_2 \right) - \theta_1 X_0^{(0)}; \Phi_2 = X_1^{(0)} \theta_1 (1-a_1^{(0)}) - X_2^{(0)} \left( \frac{\theta_3}{(s+\beta)} - \theta_2 \right).$$ [A.4c]

The coefficients $\theta_1, \theta_2$ and $\theta_3$ are defined in Eq.13a.

## Appendix B: Explicit expression for the response function $\chi_a(t)$ for small $L$

In the limit $L \to 0$ (in reality, when $L \ll 20\mu M$, see Table I) the activation probability continuously varies between 0 and 1 from $m=0$ to 4. Strictly speaking, a 3-state model cannot be justified in this case, unlike the two cases discussed in the main text. However, in this case, we may consider collapsing the states $m=0$ and 1 to a single $M=0$ state, $m=3$ and 4 into a single $M=2$ state, and the original $m=2$ will be the intermediate $M=1$ state in the reduced model, which activation probability ½. In this case, from Eq.14, a direct equation results for the perturbation in the active fraction of kinases:

$$\frac{d\delta a^*}{dt} = X_1^{(0)} \frac{d\delta a_1}{dt} + \frac{1}{2}\left[(1-a^{*(0)})\delta r - a^{*(0)}\delta b - (r^{(0)} + b^{(0)})\delta a^*\right]$$ [B.1]

Note that, in agreement with the assumption of perfect adaptation, the rate of change of the active receptor fraction depends only on the time derivative of the change in the activation probability (and therefore the attractant concentration) and not on the activation probability itself. Using Laplace transforms, Eq.B.1 is reduced to the form

$$\delta \tilde{a}^*(s) = \tilde{\chi}_a(s) \delta \tilde{a}_1(s) \text{ with } \tilde{\chi}_a(s) = \frac{X_1^{(0)} s(s+\beta)}{(s+r^{(0)}c_1 + b^{(0)}c_2)(s+\beta) - \theta_3 a^{*(0)}/2},$$ [B.2a]

where, we have defined two dimensionless constants

$$c_1 = \frac{K_r}{2[K_r + A_0(1-a^{*(0)})]}; c_2 = \frac{K_b}{2[K_b + A_0 a^{*(0)}]}.$$ [B.2b]

After explicit inversion of the Laplace transform in Eq.B.2a, we find that

$$\chi_a(t) = X_1^{(0)} \left[ \delta(t) + e^{-\theta t} \left( \beta \cosh \beta' t - \frac{\theta \beta + \theta^2 - \beta'^2}{\beta'} \sinh \beta' t \right) \right] \quad t \geq 0 \qquad [\text{B.3a}]$$

and

$$\chi_a(t) = 0 \text{ for } t < 0 \qquad [\text{B.3b}]$$

where

$$\theta = \left( \beta + r^{(0)} c_1 + b^{(0)} c_2 \right)/2 \text{ and } \beta' = \sqrt{\left( \beta + r^{(0)} c_1 + b^{(0)} c_2 \right)^2 / 4 - \lambda}. \qquad [\text{B.3c}]$$

The appearance of the Dirac $\delta$-function in Eq.B.3a may appear unphysical, but in reality, this is only an artifact of the assumption that the ligand binding and dissociation processes take place infinitely fast. Also, the experimentally measured quantity in Segall, Block and Berg (1986) is the change in the CCW-CW bias in response to a pulse stimulus, which is more similar to $\chi_R(t)$. By Eq.19, $\chi_R(t)$ is related to $\chi_a(t)$ through

$$\chi_R(t) = \frac{H^* k_Z}{a^{*(0)}} \int_0^t dt' e^{-k_Z(t-t')} \chi_a(t'). \qquad [\text{B.4}]$$

The $\delta$-function singularity in $\chi_a(t)$ is replaced by an exponentially decaying term in $\chi_R(t)$, for $t \geq 0$.

| Methylation state of receptor | Dissociation constant of ligand binding $K_m$ (mM) | Activation probability $a_m^{(0)}$ (dimensionless) |
|---|---|---|
| 0 | $27 \times 10^{-4}$ | 0.0 |
| 1 | $20 \times 10^{-3}$ | 0.25 |
| 2 | $150 \times 10^{-3}$ | 0.5 |
| 3 | $150 \times 10^{-2}$ | 0.75 |
| 4 | 60 | 1.0 |

Table I: Dissociation constant for attractant binding and activation probability at zero attractant concentration for the different methylation states of the Tar receptor [reproduced from Kollman et. al., 2005].

| Quantity | Symbol (this paper) | Experimental value |
|---|---|---|
| CheA concentration | $A_0$ | 5.3 μM |
| CheY concentration | $Y_0$ | 9.7 μM |
| CheR concentration | $R_0$ | 0.16 μM |
| CheB concentration* | $B_0$ | *2.28 μM (0.28μM) |
| CheR-CheA binding | $K_r$ | 0.39 μM |
| CheB-P – CheA binding | $K_b$ | 0.54 μM |
| Methylation time constant | $\omega_r$ | 0.75 s$^{-1}$ |
| Demethylation time constant | $\omega_b$ | 0.6 s$^{-1}$ |
| CheA auto-phosphorylation rate | $\omega_p$ | 23.5 s$^{-1}$ |
| CheY phosphorylation rate | $k'_p$ | 100 μM$^{-1}$ s$^{-1}$ |
| CheY-P dephosphorylation rate* | $k_z$ | *10 s$^{-1}$ (30 s$^{-1}$) |
| CheB phosphorylation rate | $k_p$ | 10 μM$^{-1}$ s$^{-1}$ |
| CheB-P dephosphorylation rate | $d_b$ | 1.0 s$^{-1}$ |

TABLE II: A list of the experimentally measured parameters used in this paper, from Emonet and Cluzel (2008), which also gives the original references for some of these numbers. The numbers with (*) are chosen differently for better agreement with the experimentally observed response function, as measured by Segall, Block and Berg (Segall et. al., 1986), see discussion in text. For these cases, the figures in parantheses are the values used in Emonet and Cluzel(2008).

| $L$ (mM) | $L_z$ ($\mu M / \mu m$) | $v_d$ (theoretical), $\mu m / \sec$ | $v_d$ (experimental) $\mu m / \sec$ |
|---|---|---|---|
| 0.06 | 0.08 | 2.32 | 12.6 |
| 0.06 | 0.05 | 1.45 | 9.3 |
| 0.29 | 0.15 | 0.46 | 2.7 |
| 0.5 | 0.28 | 0.85 | 3.1 |

TABLE III: Comparison of the drift velocity theoretically obtained with the experimentally measured drift velocity as measured in Ahmed and Stocker(2008).

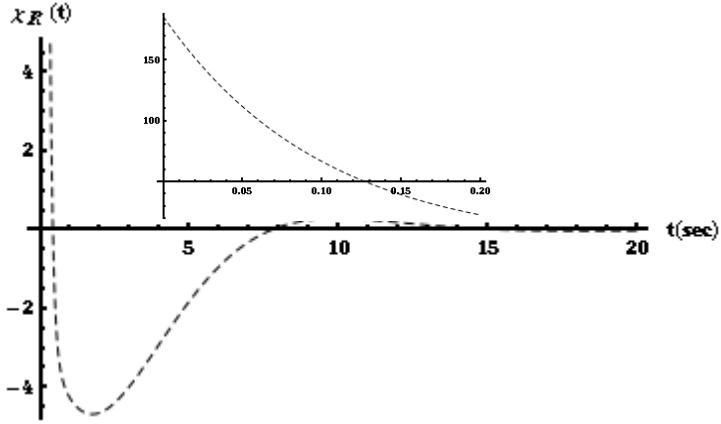

FIG 1. The chemotactic response function $\chi_R$ as a function of time, using the parameters in Table II. The inset shows that the curve has a finite intercept on the y-axis as $t \to 0$.

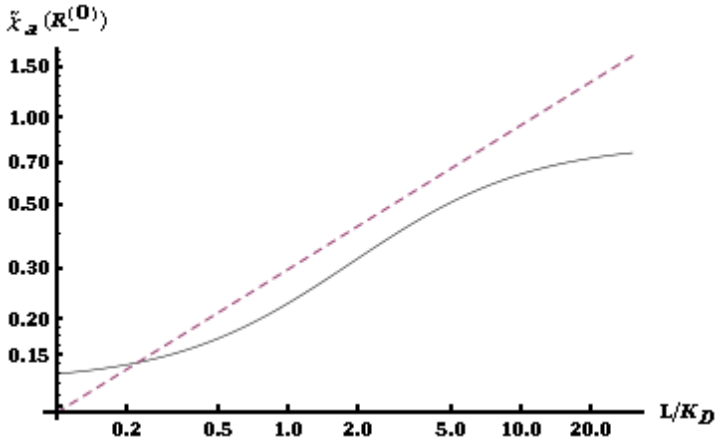

FIG.2. The Laplace transform of the response function of the active receptor fraction, $\tilde{\chi}_a(R_-^{(0)})$ as a function of the scaled attractant concentration in a log-log scale. The dashed line is a power law function, proportional to $L^{1/2}$. The sub-linear increase with $L$ in the intermediate regime indicates that, even in this regime, strictly logarithmic sensing is unlikely to be observed within the model.

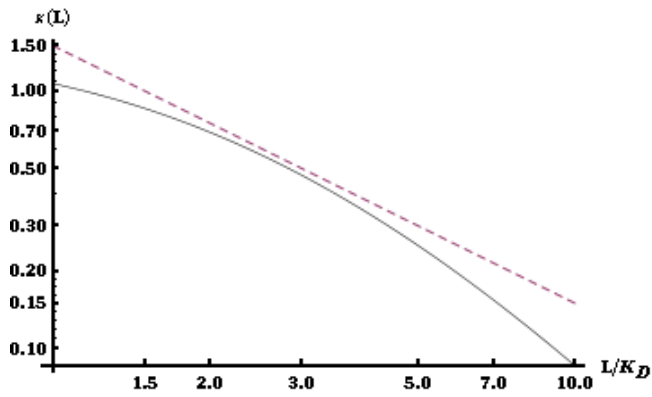

FIG.3. The figure shows the variation of the chemotactic sensitivity $\kappa(L)$ with the dimensionless attractant concentration $L/K_D$ in the range specified by Regime-II (see text) in a log-log scale. The straight line is a fit, proportional to $L^{-1}$.

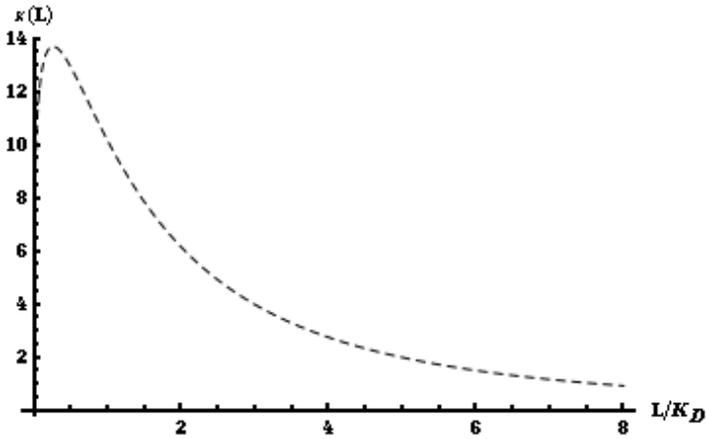

FIG.4. The figure shows the variation of the chemotactic sensitivity $\kappa(L)$ against the dimensionless attractant concentration $L/K_D$ in normal scale.

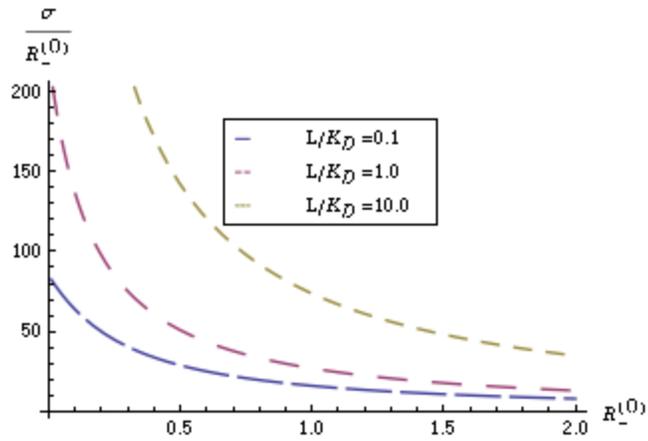

FIG 5. The figure shows the dependence of the tumble-rate dependent part of the drift velocity, as a function of the unperturbed tumble rate. The decreasing nature of the function indicates that directional persistence would increase the drift velocity, especially at high concentrations, in agreement with the more systematic study of Locsei[2007].